# Detection of Denial of Service Attacks against Domain Name System Using Neural Networks

Samaneh Rastegari, M. Iqbal Saripan and Mohd Fadlee A. Rasid

Department Of Computer and Communication System Engineering,
Universiti Putra Malaysia, Serdang, Selangor, Malaysia

## Abstract

In this paper we introduce an intrusion detection system for Denial of Service (DoS) attacks against Domain Name System (DNS). Our system architecture consists of two most important parts: a statistical preprocessor and a neural network classifier. The preprocessor extracts required statistical features in a short-time frame from traffic received by the target name server. We compared three different neural networks for detecting and classifying different types of DoS attacks. The proposed system is evaluated in a simulated network and showed that the best performed neural network is a feed-forward backpropagation with an accuracy of 99%.

*Keywords: Network security, Domain name system, Denial of service, Neural Network.*

## 1. Introduction

The Domain Name System (DNS) is one of the important elements of the Internet infrastructure which provides name to address mapping services. Attackers try to exploit this fact in order to achieve an unauthorized access and provide unavailability of communication to legitimate users. Therefore, it is very critical for network administrators to identify the variety of attack techniques in order to overcome them.

From the point of view of traffic analysis, Kambourakis *et al.* [1] proposed a practical scheme that has the ability to distinguish between authentic and falsified DNS replies. Their scheme monitors DNS traffic and based on the one-to-one mapping of DNS request and responses alerts security supervisors when the targeted name server receives responses that have not been sent out any corresponding requests previously. The main problem of their system is the rapid increase of database size in cases of high traffic rate. Wang *et al.* [2] also attempts to understand anomalous behaviors of name servers by mining DNS traffic. They focused on the query amount and the correlation between clients and servers for detecting anomalous behaviors of name servers. Their method is suffering from the lack of the potential bias in the anomaly detection and also the causes of anomalous behaviours have not been pointed.

The aim of this article is to exploit the learning ability of neural networks to detect DoS attacks against DNS. Therefore, our objectives are to generate required data for different attack scenarios and to evaluate three of the best performed neural networks for detection and classification in Intrusion Detection Systems (IDSs) which are Back-Propagation (BP), Radial Basis Function (RBF), and Self-Organizing Maps (SOM) neural networks.

## 2. DoS attacks against DNS

Two of the most common DoS attacks against DNS are direct DoS attacks and amplification attacks. For the direct DoS attacks, attacker tries to overwhelm the server by sending an excess traffic from single or multiple sources. Based on the measurement of DNS during the periods of distributed DoS attacks in [3], a huge number of query packets were received by the target name server. Therefore, the name servers flooded by DoS attacks will experience packet loss and can not always respond to every DNS request. The percentage of lost incoming DNS requests due to the excessive load is based on the DoS attack intensity [4]. Wang *et al.* [2] also points that the packet size of DNS data flow is small and the message amount is little that make the process of anomalous behaviours detection more difficult.

On the other hand, attackers establish the most sophisticated and modern type of DoS attacks known as amplification attacks to increase the effect of normal DoS attacks. According to [5], the attacks against TLD (Top-Level-Domain) name servers are all forms of DNS amplification attacks. The reason behind the name of amplification attack is that the attacker makes use of the fact that small queries can generate much larger UDP packets in response [6]. Nowadays, the attackers use the recent extension of DNS protocol (RFC 2671) to magnify the amplification factor. For example a 60 bytes DNS request can be answered with responses of over 4000 bytes. This yields an amplification factor of more than 60. Several researchers have studied the effects of reflected amplification attacks. Based on their analyzes, patterns of these attacks include a huge number of nonstandard packets larger than the standard DNS packet size which is 512 bytes [7].

## 3. Materials and Methods

This section presents a new attack detection system for DoS against DNS, which uses neural network classifier to detect and classify attacks.

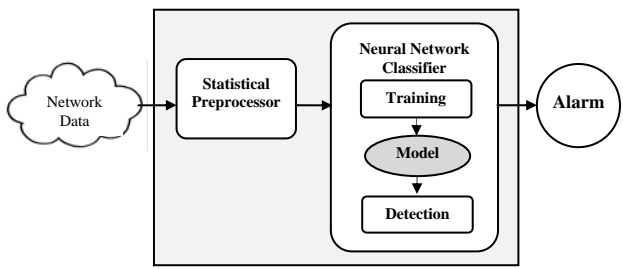

Fig. 1 System Architecture.

As indicated in Fig. 1, this system gathers network data received by a name sever. Next the preprocessor starts analyzing the traffic statistically. Finally, the neural network classifier detects abnormal activity patterns and generates an alarm in the case of attack detection. It also has the ability to report the type of DoS attack detected. The statistical preprocessor receives network traffic, and based on an administrator specified time window extracts a set of statistical variables in order to reflect the network status. In our system, we assumed a 20s length for each monitoring time slice which is more than the maximum lookup latency. The parameters that are going to characterize the DNS traffic received by the name server and that constitute the input of the classifier are defined as follows:

- Throughput of received DNS requests that is defined as the number of received bits at the server. We measured the average value of this metric for the specified time window.
- Average size of received packets by the server during a monitoring time window.
- Packet loss that is defined as the number of lost DNS packets that did not reach their destination due to the flooding attack traffic.

After preprocessing the network traffic received by the name server, a vector of 3 parameters obtained, which will be the input to the neural network classifier. This type of classifier is widely used as an effective approach to classify patterns. In this article, we investigated three types of neural networks which are BP, RBF, and SOM neural networks and compared the results to find the best performed network to be used as an IDS.

### 3.1 Experiments on BP neural network

A three layered BP neural network is evaluated in this section. The number of the units in the input layer is equal to the features of input vector which in our case is three features of DNS traffic introduced before. There are also three units in the output layer representing different states of normal and DoS attacks: [0 0 0] for normal conditions, [0 0 1] for direct DoS attack and [0 1 0] for the amplification attack.

In order to find the optimal BP classifier to fit the nature of the DNS intrusion detection data, several neural networks all different in training function and structure were trained and evaluated. At first, the best performed training function in terms of training and testing performance should be identified. In our experiments, a Levenberg-Marquardt backpropagation training function gives the acceptable training and testing errors. In the next step, different structures of BP neural networks were evaluated by varying the number of hidden neurons in the hidden layer. We should have small training error and testing error while selecting moderate number of hidden units. After tuning the BP neural network in these experiments, the optimal structure of our network with 7 neurons in the hidden layer was found. Fig. 2 & 3 show the DoS detection rate and accuracy of the system while number of hidden neurons is increasing. Performance of system during training and testing phases are presented in Fig. 4. In Fig. 5, false alarm rate of the system is presented while number of hidden neurons is changing from 3 to 21.

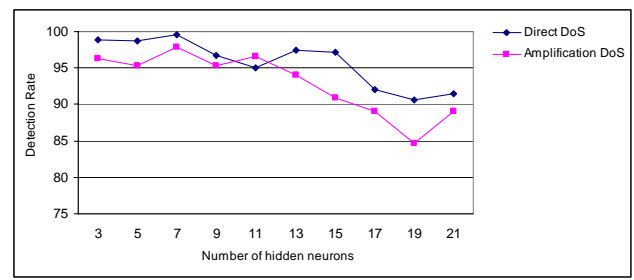

Fig. 2 DoS detection rate Vs. different number of hidden neurons.

Therefore, the main assumptions considered for training process of BP networks are listed as follows: number of epochs = 500, MSE (Mean Square Error) = 0.00001, training function = trainlm, activation function = tan-sigmoid and number of neurons in the hidden layer = 7.

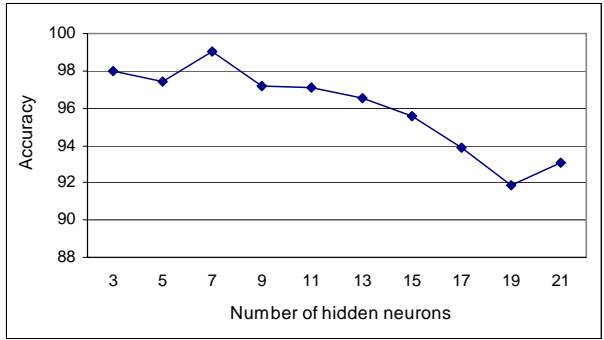

Fig 3. System accuracy Vs. different number of hidden neurons.

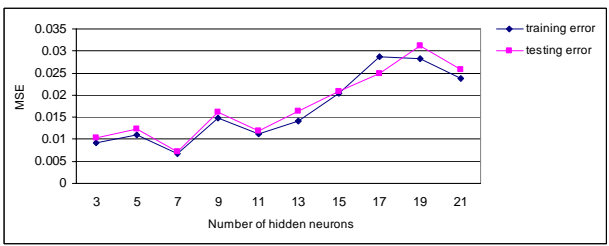

Fig 4. Mean square error of training and testing phases Vs. different number of hidden neurons.

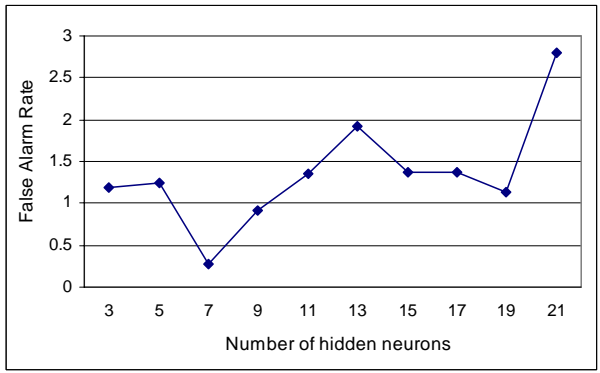

Fig 5. False alarm rate Vs. different number of hidden neurons.

## 3.2 Experiments on RBF neural network

In order to implement the RBF network for our classification problem, the activation function for the hidden units and the centers and widths of RBFs should be specified. For this purpose, the centroid locations were determined by K-mean clustering algorithm [8] and then these centers were used to calculate the width ($\sigma$) value by the following equation:

$$\sigma = \frac{\text{maximum distance between any 2 centers}}{\sqrt{\text{number of centers}}} \qquad (1)$$

These are the mean and variance of the Gaussian activation function used for the RBF's non-linear function. It is important that the width parameter be large enough that the neurons can respond to the overlapping regions of the input space, but too large values yield erroneously responses from all neurons.

Because of high calculation power requirements, it is not possible to achieve the same MSE as BP neural networks. So, the value of MSE was set to 0.001.

## 3.3 Experiments on SOM neural network

SOM is an unsupervised neural network algorithm which is mainly used for dimensionality reduction. The training is based on the competitive rules. When a learning vector is put on the input of network, its distance to weight vector is computed. The neuron with the lowest distance becomes the winner. Based on the SOM network architecture, the winner neuron and the neurons close to it are updated (winner takes all).

In this experiment, the input vector of three features has been normalized due to the large variations of input values. If the raw data is applied to the network directly, the input samples with higher values may lead to suppress the influence of smaller values. So, we used the standard normalization given by Eq. (2) was used.

$$nv[i] = v[i] \div \sqrt{\sum_x v[x]^2}$$

There are different number of neurons that were tested in order to find the best performed network. The sample results are obtained by looking at the output of the classifier applied to the trained data and noticed that all normal traffic was clustered between a specified range and the suspicious traffic was outside this cluster indicating a possible attack. After the results show a good confidence level, the trained network was evaluated by subjecting it to the test data.

In the first experiments, after the SOM network is trained using 3, 6, 25, and 40 neurons, we found that the best number of neurons for the SOM network is 25. Therefore, the SOM network parameters are considered as follows: number of epochs = 1000, number of neurons = 25, neighbors topology = Hextop, distance function = Linkdist, ordering phase learning rate = 0.9, ordering phase steps = 1000, tuning phase learning rate = 0.02 and tuning phase neighbor distance = 1.

## 3.4 Data set

Data collection is usually a basic requirement for any IDS. The type of data that we should collect is based on the type of IDS. There are two approaches for normal and intrusion traffic simulations. One is to simulate using a testbed in a real environment. The other is to simulate using simulator softwares. When accessing to a real environment for traffic simulation is hard, we exploit the power of network simulators. According to our knowledge, there are no available generated dataset for DoS attacks against DNS. Therefore, the required data for our experiments was generated using a network simulator. Our model is simulated with an OTcl program under the NS-2 (version 2.28) for different DoS attacks against DNS. The network topology of our simulation contains a single legitimate client, an attacker, and two servers. All nodes are connected to the same router. All the links are 100Mbps and 10ms except the link between target server and router that is 10Mbps and 10ms delay. A queue size of 100 packets, with a drop-tail queuing strategy was used. There are two types of traffic generated in the network which are legitimate traffic and attack traffic. A modified version of Agent/Ping with a maximum of 3 retransmissions with 5-second timeouts is used for DNS as implemented in [9]. The request interarrival period is fixed at 10s. The attacker is expected to flood the target name server with excess traffic. For this purpose, the DoS traffic is modelled as constant bit rate (CBR) source and currently can be generated by CBR traffic generator in NS-2 [10]. We chose different values of delay for applying to the attack start time in order to achieve variability. Finally, the 10-fold cross validation technique was used in order to provide a strong test to obtain reliable estimates.

## 4. Results and Discussions

In this section, the performance of each classifier in terms of detection rate, false alarm rate, and accuracy was compared. For better understanding of results comparison, we introduce these criteria:

- Accuracy, which refers to the proportion of data classified as accurate type in the total data. Accurate situations are True Positive (TP) and True Negative (TN), while false detected situations are False Positive (FP) and False Negative (FN). Accuracy of the system is calculated by the following equation:

$$Accuracy = \frac{TP + TN}{TP + TN + FP + FN} \times 100\% \qquad (3)$$

- Detection rate, which refers to the proportion of each type of DoS attack detected among all the same type of attack and is calculated by the following equation:

$$DetectionRate = \frac{TP}{TP + FN} \times 100\% \qquad (4)$$

- False Alarm Rate (FAR), which is defined as the percentage of the network traffic that is misclassified by the classifier. It can be calculated using the following equation:

$$FAR = \frac{FP}{FP + TN} \times 100\% \qquad (5)$$

Table 1 presents performance comparison of three neural network classifiers that have been implemented in this article. The results show that a BP neural network outperforms other types and gives a good system accuracy of 99% for attack detection with an acceptable false alarm rate.

Table 1: Comparison of three neural network classifiers

| Parameter / Classifier | Training time (sec) | DR (direct DoS) | DR (amplification attack) | Accuracy | FAR |
|---|---|---|---|---|---|
| BP | 11.03 | 99.55 | 97.82 | 99 | 0.28 |
| RBF | 2.89 | 99.62 | 89.48 | 95.9 | 0.23 |
| SOM | 3011.87 | 54.24 | 65.28 | 74.40 | 6.83 |

## 4. Conclusions

This paper describes two different types of DoS attacks against DNS and their patterns. Based on these patterns the required traffic data was simulated using ns-2 network simulator. We proposed a machine learning based system for detecting and classifying DoS attacks against DNS. For this purpose three different neural networks were evaluated. The results show that a three layered BP neural network with a 3-7-3 structure can give us 99% accuracy and a good classification rate for direct DoS and amplification attacks comparing to RBF and SOM neural networks.

**Samaneh Rastegari** received her B.S. degree from the Department of Computer Engineering at Yazd University, in 2005. Currently, she is pursuing her M.S. degree in the Department of Computer and Communication System Engineering at the Universiti Putra Malaysia (UPM). Her research areas include network security, information security, quality of service and protocols.

**M. Iqbal Bin Saripan** received the B.Eng. in Electronics Engineering from Universiti Teknologi Malaysia (UTM), in 2001, and Ph.D. in Digital Image Processing from University of Surrey in 2006. He is a senior lecturer at the Department of Computer and Communication System Engineering, Universiti Putra Malaysia (UPM). His research interests are computer, biomedical engineering, signal processing and artificial intelligence.

**Mohd Fadlee A. Rasid** is the deputy director for National Centre of Excellence on Sensor Technology (NEST) at Universiti Putra Malaysia. He received a B.Sc. in electrical engineering from Purdue University, USA and a Ph.D. in electronic and electrical engineering (mobile communication) from Loughborough University, U.K. He directs research activities within the Wireless Sensor Network (WSN) group and his work on wireless medical sensors is gaining importance in health care applications involving mobile telemedicine and has had worldwide publicity, including BBC news. He involves with UK Education and Research Initiative under the British Council on wireless medical sensors project that will allow a more patient-driven health service in improving the efficiency of health care delivery. He is also part of the European Union (EU) STIC Asia Project on ICT-ADI: Toward a human-friendly assistive environment for Aging, Disability & Independence. He currently leads a few research projects on WSN, particularly for medical and agriculture applications. He was nominated for IEE J A Lodge Award for outstanding work in Field of Medical Engineering, London, 2005 and was the proud recipient of State of Selangor Young Scientist Award in 2006.